\documentclass[twocolumn,preprintnumbers,amsmath,amssymb,amsfonts,nofootinbib]{revtex4}
\usepackage{epsfig}
\usepackage{graphicx}

\begin{document}
\preprint{}

\title{Is the Cosmological Coincidence a Problem?}

\author{Navin Sivanandam}
\email{navin.sivanandam@gmail.com}
\affiliation{African Institute for Mathematical Sciences, Muizenberg, Cape Town, SA}

\date{\today}

\begin{abstract}
The matching of our epoch of existence with the approximate equality of dark energy and dark matter energy densities is an apparent further fine-tuning, beyond the already troubling 120 orders of magnitude that separate dark energy from the Planck scale. In this paper I will argue that the coincidence is not a fine-tuning problem, but instead an artifact of anthropic selection. Rather than assuming measurements are equally likely in all epochs, one should insist that measurements of a quantity be typical amongst all such measurements. As a consequence, particular observations will reflect the epoch in which they are most easily made. In the specific case of cosmology, most measurements of dark energy and dark matter will done during an epoch when large numbers of linear modes are available to observers, so we should not be surprised at living at such a time. This is made precise in a particular model for the probability distribution for $r=\textrm{min}\left(\frac{\Omega_m}{\Omega_{\Lambda}},\frac{\Omega_{\Lambda}}{\Omega_m}\right)$, where it is shown that if $p(r)\sim \left[N(r)\right]^b$ (where $N(r)$ is the number of linear modes, and $b$ is some arbitrary positive power), the probability that $r$ is greater than its observed value of 0.4, is close to 1. Thus the cosmological coincidence is no longer problematic.
\end{abstract}

\maketitle

\section{Introduction}
\subsection*{The Coincidence Problem}
We live during a particularly interesting cosmological epoch; not only are our skies filled with the riches of the early universe, we are also fortunate enough to be alive just as the era of matter gives way to that of dark energy. Somewhat more prosaically, we observe that the fractional densities of matter and dark energy are about the same: $\Omega_m\sim\Omega_{\Lambda}$ (the $\Omega_i$ are defined as $\Omega_i=\rho_i/\rho_{\textrm{crit}}$, where $i$ denotes the particular component of the universe and $\rho_{\textrm{crit}}$ is the time-dependent critical density). This ``coincidence problem'' is often viewed as a challenge for models of dark energy, and even if not quite as troubling as the problem of the magnitude of dark energy, explaining the coincidence in question is far from straightforward \cite{Carroll:2000fy,Carroll:2001xs,Weinberg:2000yb}.

\begin{figure}
\centering
\includegraphics[width=\columnwidth]{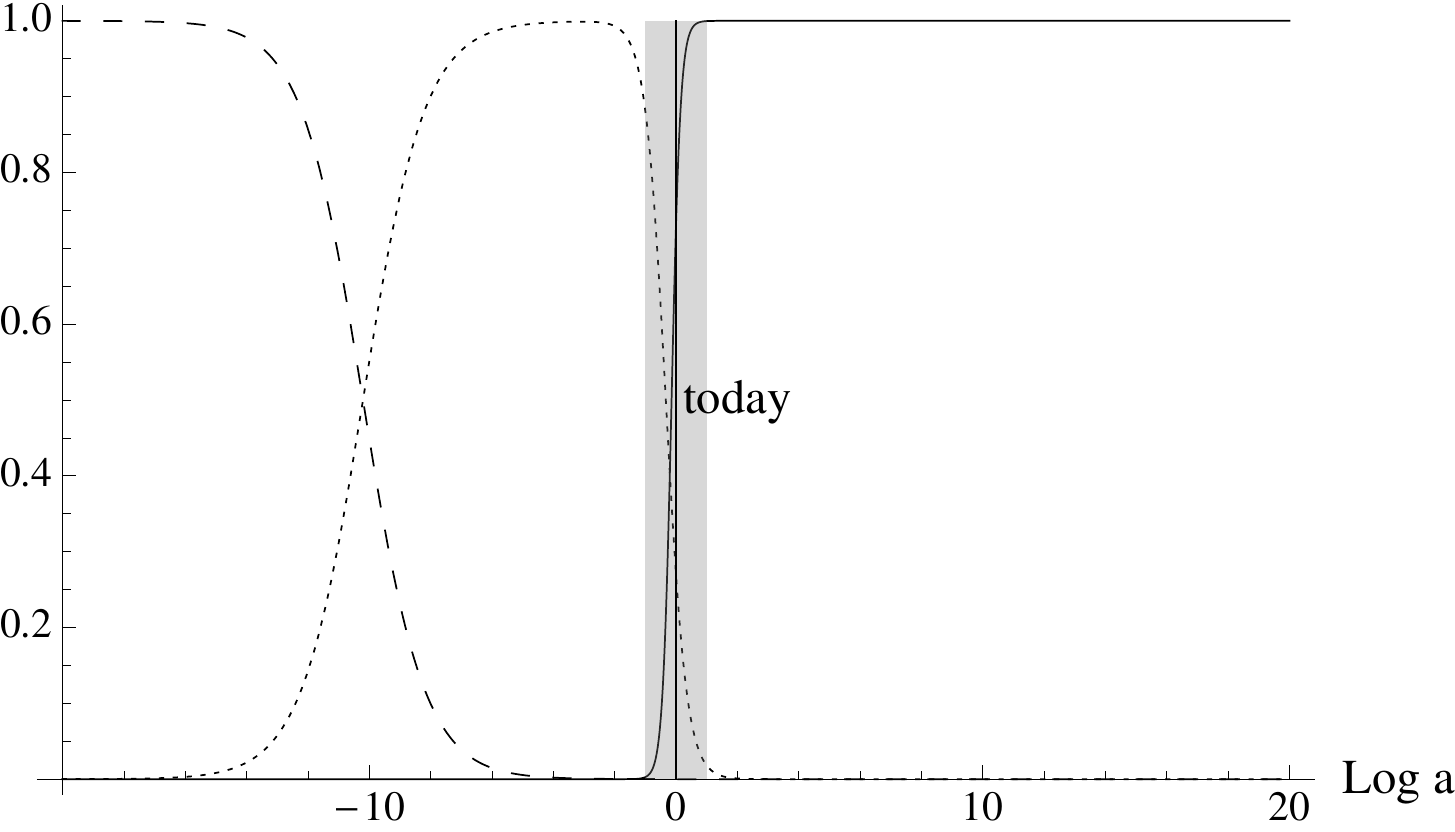}
\caption{The evolution of the different components of the universe, as a function of $\textrm{log}\,a$. The dashed line denotes $\Omega_r$, the dotted line $\Omega_m$ and the solid line $\Omega_{\Lambda}$. The shaded grey region covers the short epoch (that we happen to live in) where $\Omega_{\Lambda}$ and $\Omega_m$ are comparable in magnitude.}
\label{omegas}
\end{figure}

\begin{figure*}
\centering
\includegraphics[width=\textwidth]{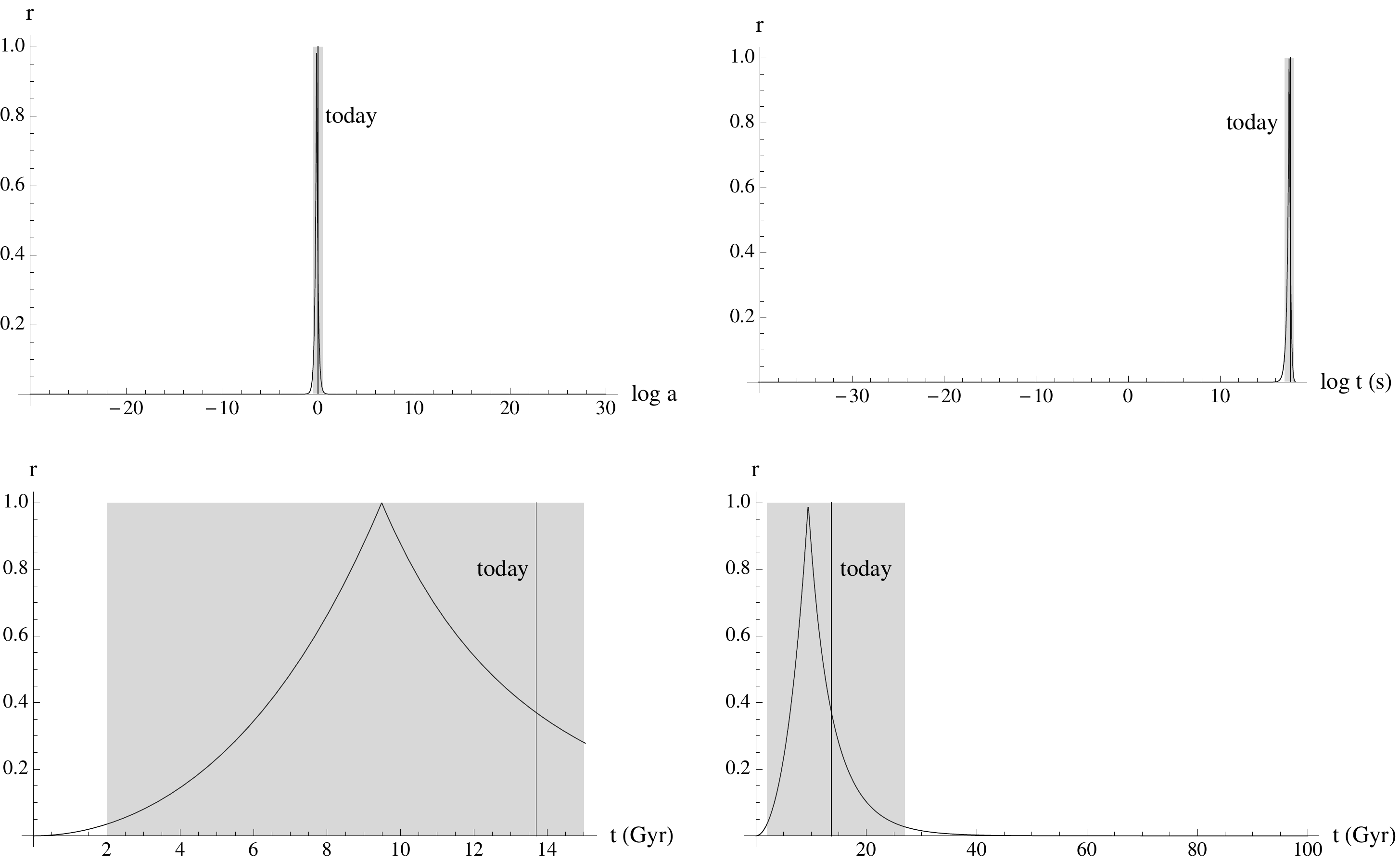}
\caption{The four plots (following those in \cite{Lineweaver:2007qh}) show the the ratio $r=\textrm{min}\left[\frac{\Omega_{\Lambda}}{\Omega_m},\frac{\Omega_m}{\Omega_{\Lambda}}\right]$ as a function of time. In each case the vertical line denotes today and the shaded grey area marks the region where $r\gtrsim 0.05$. Clockwise from the upper left the x axis shows logarithmic scale factor, logarithmic time, linear time up to today and linear time until 100 Gyr. Note that one's impression of a coincidence depends strongly on the choice of time coordinate and one's cutoffs.}
\label{rs}
\end{figure*}

The apparent problem is starkly evident in figure \ref{omegas}, which shows the evolutions of the fractional densities of the various components of the cosmological fluid as functions of the logarithm of the scale factor $a$. Notice that only in an uncomfortably narrow band of $\log a$ are $\Omega_m$ and $\Omega_{\Lambda}$ comparable. To make this discomfiture more precise I am going to follow Lineweaver and Egan in \cite{Lineweaver:2007qh} and define the useful parameter, $r$:
\begin{equation}
r=\textrm{min}\left[\frac{\Omega_{\Lambda}}{\Omega_m},\frac{\Omega_m}{\Omega_{\Lambda}}\right]\, .
\end{equation}
The current value of $r$ is around 0.4, and the coincidence problem can be rephrased as a question about the probability of finding $r\gtrsim0.4$. As Lineweaver and Egan point out, the expected value of $r$ depends on one's prior for $p(a)$, the probability distribution for when one expects to live. Figure \ref{rs} (where I have replicated similar plots from  \cite{Lineweaver:2007qh}) provides a stark visual representation of this prior dependence. The miracle of ``Why now?'' is rather less dramatic if the likelihood of our observation is flat in linear rather than log time, although the problem is still significant if we speculate on why we are not living in the far future. Regardless of the choice of prior, an examination of the shape of the plots makes it clear that the coincidence problem results from the fact that $r$ approaches zero at large and small $a$, and it is at one or both of these values that most of the probability lies, assuming that one takes a flat prior for the $x$-axis position that we happen to live at. If one assumes $p(\ln a)$ or $p(a)$ is flat, the probability of measuring  $r\sim1$ depends on time cutoffs (the details of these calculations can be found in appendix \ref{withoutselection}):
\begin{align*}
P(r\geq0.4)&\sim e^{-Ht_f}\, , & p(a)&\sim 1\\
P(r\geq0.4)&\sim\frac{1}{Ht_f-\frac{2}{3}\ln\frac{t_i}{t_1}}\, ,& p(\ln a)&\sim 1
\end{align*}
As $t_f\rightarrow\infty$ both of the above probabilities approach 0, and in the the case of $p(\ln a)\sim1$ the same is true as $t_i\rightarrow 0$.

As with many other deep issues in cosmology there is no shortage of possible solutions to the coincidence problem. These solutions can be usefully divided into two categories: those which change the dynamics of the universe and those which change our prior for $p(a)$. For the former class we have a host of dynamical dark energy models (for a review of the taxonomy of such models see \cite{Copeland:2006wr}), some of which purport to solve the coincidence problem, e.g. \cite{Dodelson:2001fq}. Solutions that change the prior for $p(a)$ are ``selection effect'' or ``anthropic'' explanations, e.g.  \cite{Weinberg:2000yb,Garriga:1999hu,Garriga:2000cv,Bludman:2000pv,Lineweaver:2007qh,Egan:2007ht}, and will be the focus of this paper.

\subsection*{Selection Effects and Anthropics}
While the search for solutions to fine-tuning/coincidence problems has been a powerful heuristic in the progress of physics, one should should remember that apparently unlikely events are often not explained by new dynamics. As Weinberg notes in \cite{Weinberg:2000yb} one striking historical example of this is the distances of the planets from the sun; these follow no particular pattern\footnote{Bode's Law excepted.} and the fine-tuning of conditions on Earth for life is readily explained by conditioning on the existence of life. Similar anthropic arguments have proved fruitful in many cosmological settings, most notably with regards to the value of the cosmological constant \cite{Weinberg:1987dv,Martel:1997vi}.

The term ``the anthropic principle'' \cite{Carter:1974zz} has come to encompass a large number of related concepts, all of which share the notion that observations necessarily require observers. The refinement of anthropic notions most used by cosmologists is Vilenkin's ``Principle of Mediocrity'' \cite{Vilenkin:1994ua,Vilenkin:2011dd}, which can be readily phrased as a statement about typicality:
\begin{quote}
``We should observe a universe that is typical amongst those containing observers.''
\end{quote}

As noted above, anthropic arguments have been applied to the coincidence problem by a number of authors. For example, in \cite{Garriga:1999hu} Garriga et al. point out that the coincidence between $t_0$ (today) and $t_{\Lambda}$ (the time of dark energy dominance) can be explained by assuming that the number of observers is proportional to the amount of carbon, so that most observations of the universe should take place at the time of peak carbon production, $t_0\sim t_{\textrm{carbon}}$\footnote{This, of course, makes the reasonable assumption that on cosmological scales the timescale of intelligent life evolving from available carbon can be ignored}. Noting that carbon peaks with star formation ($t_{SFR}\sim t_{\textrm{carbon}}$), the authors find:
\begin{equation}
t_{\Lambda}\sim t_G\sim t_{SFR}\sim t_{\textrm{carbon}}\sim t_0\, .
\end{equation}
$t_G$ is the time of galaxy formation, and the first two approximate equalities follow from anthropic (and other) details of structure formation. Thus the coincidence $t_{\Lambda}\sim t_0$ is explained.

A complementary analysis has been carried by Lineweaver and Egan \cite{Lineweaver:2007qh}. Here the authors consider the age distribution of terrestrial planets in the universe, imposing an additional offset ($\Delta t_{obs}=4\,\textrm{Gyr}$) to account for the delay between forming a planet and life evolving. Within this framework they find that 68\% of observers emerge earlier than us, while 32\% emerge later. Thus we are typical amongst observers on terrestrial planets who take around 4 billion years to evolve (the result is shown to be robust for $\Delta t_{obs}\leq10\,\textrm{Gyr}$). The argument is extended in \cite{Egan:2007ht} to apply to dynamical dark energy models, with similar conclusions.

Although the arguments above are perfectly satisfactory anthropic explanations for the coincidence problem, they are not without problems:
\begin{itemize}
\item Carbon bias. Carbon-based, planet-bound life may only be a small (atypical) subset of potential observers.
\item Sensitivity to late time observers. If the typical timescale for intelligent life to form is much greater than that of carbon production or of terrestrial planet formation, then the above methods are both missing the majority of observers.
\item What about the multiverse? If we take the multiverse seriously, we should really be analyzing all parts of it with a suitably small dark energy component, and not fixing the detailed astrophysics.
\end{itemize}

One way to assuage the these concerns is to be more general in imposing selection effects. This can be done by focussing on the particular observation being made. To rephrase the ``Principle of Mediocrity'':
\begin{quote}
``A measurement of a quantity should be typical of all possible measurements of that quantity.''
\end{quote}
To unpack that a little, consider the ratio $r$ defined above; when we ask that $r$ not be finely tuned, we are really asking that the value of $r$ that we measure be typical amongst all possible values that could possibly be measured. Now, if the number of measurements is independent of time, we end up with the coincidence problem outlined above. If, however, measurements are easier in some epoch and harder in another and impossible in a third, we have to take that into account when asking what a typical measurement of $r$ is.

Measuring $r$ requires measuring both $\Omega_m$ and $\Omega_{\Lambda}$, so our distribution for typical values of $r$ should account for how easy it is to measure these quantities. As I shall argue below, this amounts to relating the prior distribution of $r$ to the ease of measuring the expansion of the universe. Essentially, in a universe with an accelerating component and with a decoupling scale (below which matter no longer follows the Hubble flow), there is a finite time when modes are available to do cosmology, and the number of modes available is peaked near the epoch of matter $\Lambda$ equality. As a result the probability of $\Omega_{\Lambda}\sim\Omega_m$ is close to 1.

\subsection*{The Measure Problem}
I have so far left unmentioned the measure problem (see \cite{Freivogel:2011eg} for an up-to-date overview of measures). This oversight will be extended throughout most of the paper, but it would be remiss of me not to spend a little time on the issue.

The measure problem arises in eternal inflation because sensitivity of predictions to the choice of measure on the populated landscape of possible vacua. This sensitivity can have profound consequences for ones choice of prior for where and when observers should expect to find themselves in the multiverse. Examples of measure-based solutions to the coincidence problem can be found in \cite{Bousso:2010im,Bousso:2010zi}.

While in this paper I am focussing on an anthropic approach, it is important to note that )as with all discussions of fine tuning is cosmology) considerations of the measure may also be relevant and could affect my conclusions. That said, a measure-blind approach to issues of selection and fine-tuning in cosmology has not been without its successes in the past \cite{Weinberg:1987dv}, and the present work hopes to follow in those footsteps.

\subsection*{Organization}

I shall provide the details of my argument that the number of modes provides a good proxy for $p(r)$ in section \ref{observed}, following this with a derivation of the appropriate probability distribution and a calculation of $P(r\geq0.4)$ under various different assumptions. After this, in section \ref{conclusions}, I will discuss the conclusions one might draw from this sort of reasoning, along with a myriad of caveats and qualifications.

\section{Measuring $r$}\label{observed}
To make the above discussion on measuring $r$ more precise, let us begin considering how a general observer may go about measuring $\Omega_{\Lambda}$ and $\Omega_m$. Because matter clumps, its presence  can be detected through the motions of luminous test particles moving in the potential well of a particular clump of matter. As such, sufficiently clever cosmologists armed with sufficiently advanced instruments and sufficiently generous funding grants can locate and ``weigh'' clumps of matter. Then, by adding the masses of these clumps together the aforementioned cosmologists can make reasonable estimates of $\Omega_m$ during almost all cosmological epochs (with the obvious caveat that this argument requires luminous matter to roughly track dark matter).

Measuring $\Omega_{\Lambda}$, however, is a different story. Assuming there no significant spatial variations in the dark energy density, $\Omega_{\Lambda}$ can only be detected through measurements of the  Hubble expansion. This expansion is detected through observations of the redshifts and distances of ``objects'' that are not gravitationally bound to the observer. The greater the number of such objects, the easier it is to measure the expansion. Although ``number of objects'' is inherently a notion that depends on the cosmological and astrophysical details of the particular universe we find ourselves in, one has a reasonable proxy in the number of linear modes (modes corresponding to scales that have not undergone gravitational collapse) in the observer's Hubble radius that are larger than the largest gravitationally bound (and thus decoupled from the Hubble flow) structure.

There are several reasons why the use of $N$ (the number of linear modes) is a good proxy for the number of objects. Firstly, it bounds the maximum number of independent (in the sense of the motion with the Hubble flow) objects. As well as the number of modes, $N$ also counts the number of volumes within a Hubble radius that can contain a single (at most) maximally-sized bound structure; thus $N$ bounds the maximum number of such independent objects. Of course, a) ``maximum'' is not the same as ``number of''  and b) each of these independent volumes may contain many objects (for example, SNIa) which can be used to measure distance. That said, so long as the universe is isotropic it is reasonable to assume that the number of objects should scale as $N$.

In addition to the above line of reasoning, we should also note that $N$ directly characterizes our ability to measure cosmological parameters when we use the linear modes themselves as probes, as with the CMB (and, in the future, with 21cm observations). Although such measurements usually cannot constrain dark energy by themselves (see \cite{Sherwin:2011gv} for an example of a CMB-only constraint on dark energy), they are an important part of our ability to accurately determine cosmological parameters, including $r$.

Finally, assuming an approximately scale-invariant spectrum (as one has in our Hubble volume and would have in other Hubble volumes with an inflationary period in their past) also implies a correlation between number of linear modes and objects. This follows since with a scale-invariant spectrum the initial amount of power at each scale is constant. Consequently the number of small objects useful for probing cosmology should scale with $N$.

There are, of course, many other factors that will affect the ease of measuring $r$. However, on the ground of maintaining generality, I am going to ignore most of them. One that might have a general applicability, though, is the ability to discriminate between different cosmologies. In particular the ability to tell an accelerating from non-accelerating universe seems a prerequisite for measuring $r$, and this is not independent of the value of $r$. This will be discussed in more detail below.

In order to encompass a wider class of models than simply vacuum energy, I shall present results for a cosmology consisting of matter with the equation of state $p=0$, and dark energy with the equation of state $p=w\rho$, where $-1\leq w<-1/3$. Then:
\begin{equation}
H^2=\frac{1}{3}\left(\rho_{m0}\left(\frac{a_0}{a}\right)^{3}+\rho_{\Lambda 0}\left(\frac{a_0}{a}\right)^{3+3w}\right)\, .
\end{equation}
Here, as below, the reduced Planck mass, $(8\pi G)^{-1/2}$ has been set equal to 1 and the subscript $0$ denotes the time when the largest bound structure for a given observer enters the Hubble radius. I have also assumed that we are in a flat universe with $\Omega_{rad}\ll1$.\footnote{It would take a particularly delicate fine-tuning to arrange a period of radiation domination close to the time of matter/dark energy domination}

I should emphasize that while the rest of this section is littered with details of calculations, multiple plots and several actual numbers, these are somewhat incidental to the larger argument. The purpose of this paper is not to claim that the probability of measuring $r$ has some value that can be calculated given a suitably detailed model of physicists and the methods of observation. Rather, I wish to point out that the apparent fine-tuning of the coincidence problem is an artifact of an error in the typical choice of prior for the epoch in which cosmological observers live. Once this error is corrected, by noting that an appropriate prior for the epoch of measurement will take into account the ease (or otherwise) of making the measurement, one finds that the fine-tuning has vanished. The particular analysis and numerical results that follow should thus be considered as evidence for this point of view, and a representative (rather than faithful) model of reality.

\subsection{Number of Measurements}
The number of independent measurements that can be made of the expansion of the universe depends on the number of modes within a Hubble radius that are not decoupled from the Hubble flow. In a universe of matter and dark energy, modes enter the Hubble volume of an observer during the epoch of matter domination and exit during the epoch of dark energy domination, thus there is only a finite period during which cosmological measurements can be made.

\begin{figure*}
\centering
\includegraphics[width=0.9\textwidth]{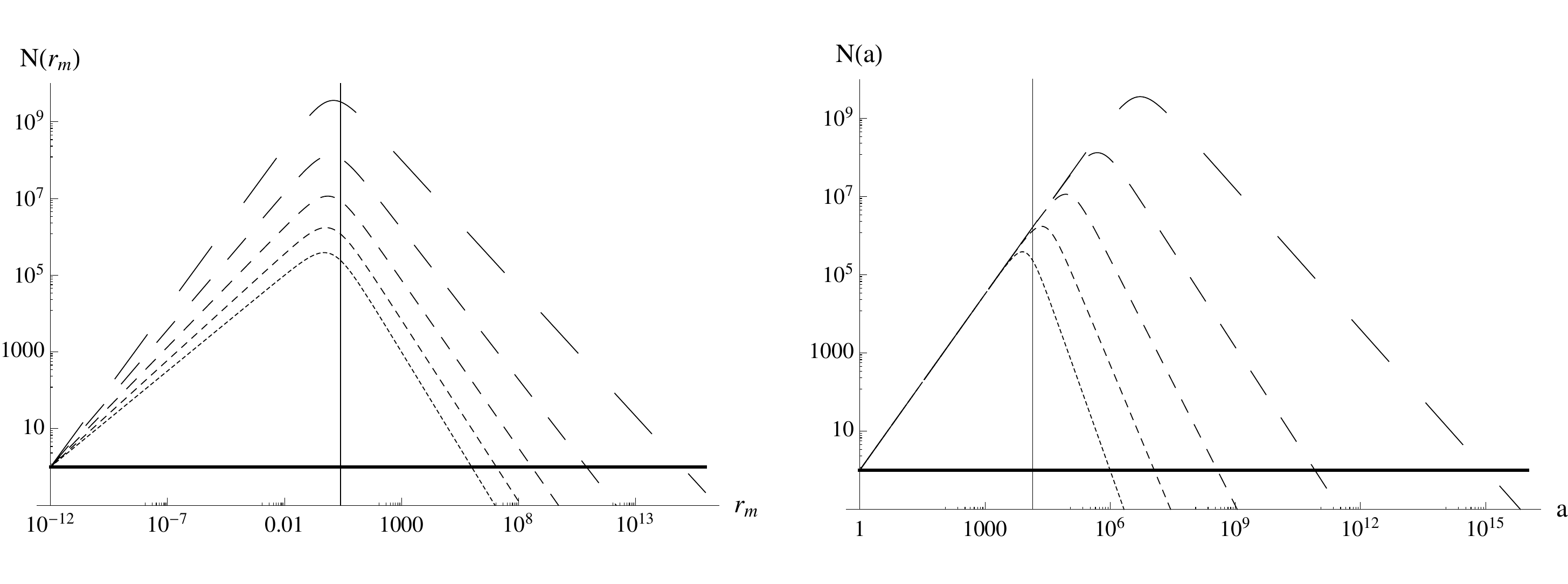}
\caption{These two plots illustrate how the number of modes available to an observer changes as a function of $r_m$ and of $a$. In each case $r_0=10^{-12}$ (which is approximately the value for our universe, if we assume dark energy is a pure cosmological constant). $w=-1$ for the most finely dashed line and then -0.9, -0.8, -0.7 and -0.6 as the dashing increases in width. Note the increasing $w$ increases the amount of time available for cosmology but has little effect on the shape (and importantly the peakedness) of $N(r_m)$. In each case the vertical line indicates the value of $r_m$ or $a$ for current observers in our universe. }
\label{plotnnra}
\end{figure*}

Because the definition of $r$ varies with time it is useful to work with the quantities $r_m=\Omega_{\Lambda}/\Omega_m$ and $r_{\Lambda}=\Omega_m/\Omega_{\Lambda}$, which are equal to $r=\textrm{min}\left[\frac{\Omega_{\Lambda}}{\Omega_m},\frac{\Omega_m}{\Omega_{\Lambda}}\right]$ when $\Omega_{\Lambda}<\Omega_m$ or $\Omega_m<\Omega_{\Lambda}$ respectively. With these definitions, we have:
\begin{align}
H^2&=\frac{\rho_{\Lambda}}{3}\left(\frac{1}{r_m}+1\right)\, ,\nonumber\\
\rho_{\Lambda}&=\rho_{\Lambda0}\left(\frac{a_0}{a}\right)^{3+3w}\, ,\nonumber\\
r_m&=r_0\left(\frac{a_0}{a}\right)^{3w}\, .
\end{align}
Let $k_*$ be the comoving wavenumber corresponding to the largest bound structure for a given observer, and consider the quantity $k_*/aH$:
\begin{equation}
\frac{k_*}{aH}=\frac{k_*}{a_0H_0}\left(\frac{1+r_0}{r_0^{-\frac{1}{3w}}}\right)^{\frac{1}{2}}\left(\frac{r_m^{-\frac{1}{3w}}}{1+r_m}\right)^{\frac{1}{2}}\, .
\end{equation}
If we set subscript 0 quantities as the initial time when the mode corresponding to the largest bound structure is equal in size to the Hubble radius, then we have:
\begin{equation}\label{kaH}
\frac{k_*}{aH}=\left(\frac{1+r_0}{r_0^{-\frac{1}{3w}}}\right)^{\frac{1}{2}}\left(\frac{r_m^{-\frac{1}{3w}}}{1+r_m}\right)^{\frac{1}{2}}\, .
\end{equation}
Cosmology is possible when the above quantity is larger than 1. This is true, in the case of $w=-1$, when:
\begin{equation}
r_0<r_m<\frac{(r_0+1)\sqrt{r_0^2+4r_0}-r_0^2-3r_0}{2r_0}\, .
\end{equation}
This range is somewhat larger for larger values of $w$, as can be seen in figure \ref{plotnnra}. The number of modes available for cosmology scales like the cube of (\ref{kaH}):
\begin{equation}
N_m\sim\left(\frac{1+r_0}{r_0^{-\frac{1}{3w}}}\right)^{\frac{3}{2}}\left(\frac{r_m^{-\frac{1}{3w}}}{1+r_m}\right)^{\frac{3}{2}}\, .
\end{equation}
Figure \ref{plotnnra} shows the number of modes available to an observer as a function of $r_m$ and of $a$; note that while changing $w$ affects the length of the period during which cosmology can be done, the shape of the distribution of modes as a function of $r$ or $a$ is always strongly peaked and broadly unchanged for different values of $w$.

In order to calculate the probability $P(r\geq0.4)$, we will need the number of modes in terms of $r_{\Lambda}$ when $\Omega_{\Lambda}>\Omega_m$. This is given by:
\begin{equation}
N_{\Lambda}\sim\left(\frac{1+r_0}{r_0^{-\frac{1}{3w}}}\right)^{\frac{3}{2}}\left(\frac{r_{\Lambda}^{1+\frac{1}{3w}}}{1+r_{\Lambda}}\right)^{\frac{3}{2}}\, .
\end{equation}

\subsection{Probability Distribution for $r$}
To calculate the probability distribution function $p(r)$, we begin by constructing $p(r|r\equiv r_m)$ and $p(r|r\equiv r_{\Lambda})$. The probabilities should be proportional to a function of the number modes available:
\begin{align}\label{prf}
p(r|r\equiv r_m)&\sim f_m\left(N_m(r)\right)\nonumber \\
p(r|r\equiv r_{\Lambda})&\sim f_{\Lambda}\left(N_{\Lambda}(r)\right)\, .
\end{align}

There are at several reasonable choices for the form of the function $f_i$, depending on one's model of measurement. On general grounds, one should expect the $f_i$ to be monotonic (measurement is clearly easier when more modes are available), but beyond that it is hard to justify a particular choice of $f_i$. For definiteness I will begin with the $f_i$ as powers of $N_i$, this corresponds to the assumption that an observer's awareness of the value of $r$ scales with the number of data points available. Another plausible choice of model of the measurement process would be to assume that measurements are impossible below some threshold in $N$, but equally likely above said threshold\footnote{I am grateful to Mike Salem for raising this issue.}. This choice is covered in appendix \ref{threshold}.

In the case when the $f_i$ are powers of $N$:
\begin{align}
p(r|r\equiv r_m)&=A_m\left[N_m(r)\right]^{b_m}\nonumber \\
p(r|r\equiv r_{\Lambda})&=A_{\Lambda}\left[N_{\Lambda}(r)\right]^{b_{\Lambda}}\, .
\end{align}
The quantities $A_m$, $A_{\Lambda}$ are given by the requirements that:
\begin{align}
\int_0^1 p(r|r\equiv r_m\textrm{d}r)=\int_0^1 p(r|r\equiv r_{\Lambda})\textrm{d}r=1\, .
\end{align}
One then obtains the following expressions for the conditional probabilities:
\begin{align}
p(r|r\equiv r_m)&=A_m \left(\frac{r_m^{-\frac{1}{3w}}}{1+r_m}\right)^{\frac{3b_m}{2}}\nonumber \\
p(r|r\equiv r_{\Lambda})&=A_{\Lambda} \left(\frac{r_{\Lambda}^{1+\frac{1}{3w}}}{1+r_{\Lambda}}\right)^{\frac{3b_{\Lambda}}{2}}\, .
\end{align}
The pre-factors depending on $r_0$ have been reabsorbed into the $A_i$, which are now given by:
%\begin{widetext}
\begin{align}
A_m&=\frac{2-\frac{b_m}{w}}{2\, {}_2F_1\left(\frac{3b_m}{2},\left(1-\frac{b_m}{2w}\right),\left(2-\frac{b_m}{2w}\right),-1\right)}\, ,\nonumber\\
A_{\Lambda}&=\frac{2+3b_{\Lambda}+\frac{b_{\Lambda}}{w}}{ 2{}_2F_1\left(\frac{3b_{\Lambda}}{2},\left(1+\frac{b_{\Lambda}}{2}\left(3+\frac{1}{w}\right)\right),\left(2+\frac{b_{\Lambda}}{2}\left(3+\frac{1}{w}\right)\right),-1\right)}\, .
\end{align}
%\end{widetext}
${}_2F_1$ is a hypergeometric function.

\begin{figure*}
\centering
\includegraphics[width=\textwidth]{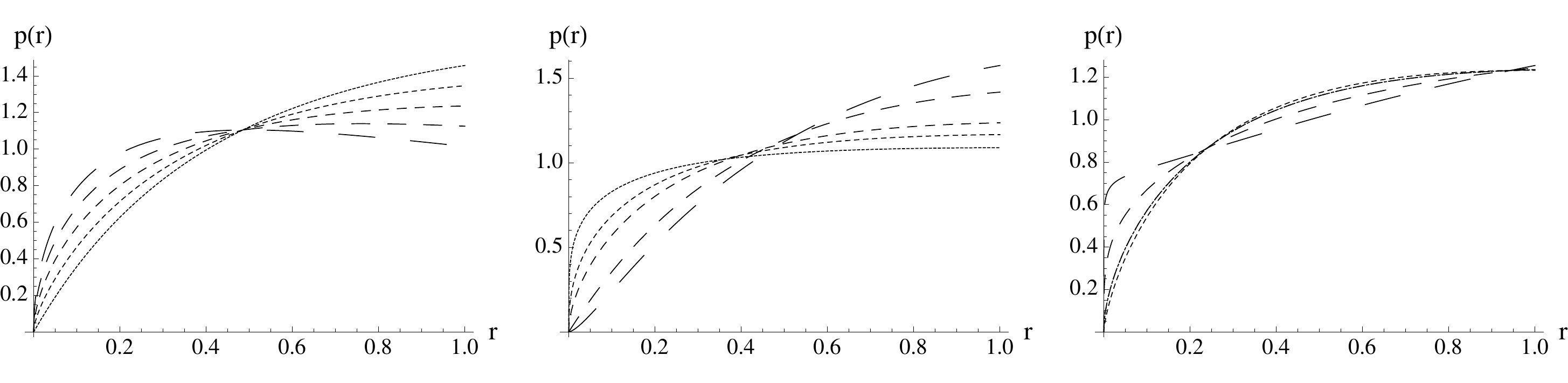}
\caption{The left-hand plot shows $p(r)$ for various values of $q$ ($b=1$, $w=-1$): in order of increasing dash width $q=0,1/4,1/2,3/4,1$. In the middle the curves are for different values of $b$ ($q=1/2$, $w=-1$): in order of dash width $b=1/3,2/3,1,2,3$.  And on the right we have varying $w$: in order of dash width we have $w=-1,-0.75,-0.5,-0.4,-0.35$.}
\label{plotprqb}
\end{figure*}

To calculate $p(r)$ from the conditional probabilities now just requires an application of Bayes' theorem:
\begin{align}
p(r)&=p(r|r\equiv r_m)P(r\equiv r_m)+p(r|r\equiv r_{\Lambda})P(r\equiv r_{\Lambda})\nonumber\\
&=q p(r|r\equiv r_m)+(1-q)p(r|r\equiv r_{\Lambda})\nonumber\\
\end{align}
In the second line I've set $q$ as the probability that we live in an epoch of matter domination. If we further assume $b_m=b_{\Lambda}$, the complete (and unwieldy) expression for $p(r)$ is then:
\begin{widetext}
\begin{multline}
p(r)=\frac{q\left(2-\frac{b}{w}\right)}{2\, {}_2F_1\left(\frac{3b}{2},\left(1-\frac{b}{2w}\right),\left(2-\frac{b}{2w}\right),-1\right)}\left(\frac{r^{-\frac{1}{3w}}}{1+r}\right)^{\frac{3b}{2}}\\
+\frac{(1-q)\left(2+3b+\frac{b}{w}\right)}{ 2{}_2F_1\left(\frac{3b}{2},\left(1+\frac{b}{2}\left(3+\frac{1}{w}\right)\right),\left(2+\frac{b}{2}\left(3+\frac{1}{w}\right)\right),-1\right)}\left(\frac{r^{1+\frac{1}{3w}}}{1+r}\right)^{\frac{3b}{2}}\, .
\end{multline}
\end{widetext}
$p(r)$ is plotted in figure \ref{plotprqb} for various values of $q$, $b$ and $w$.

\begin{figure*}
\centering
\includegraphics[width=\textwidth]{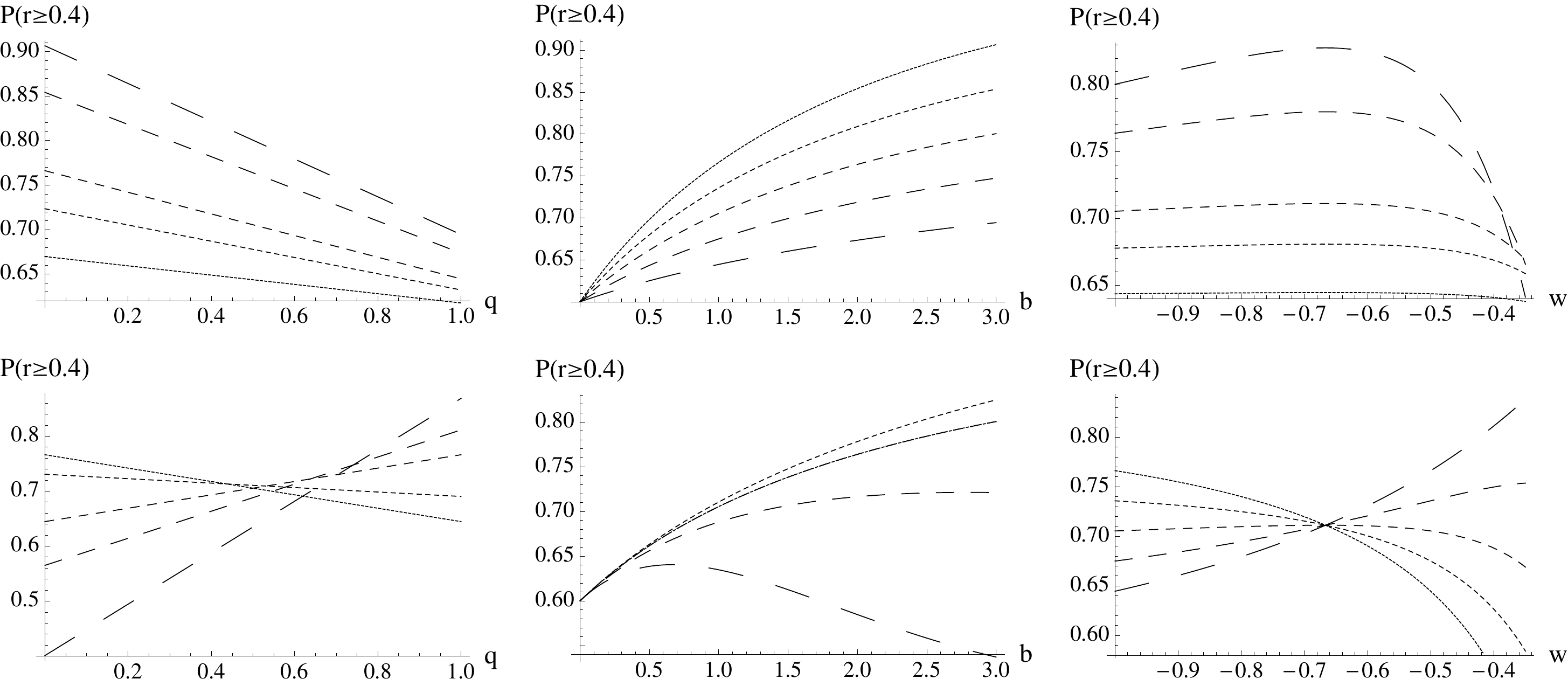}
\caption{Plots of $P(r\geq0.4)$ as functions (from left to right) of $q$, $b$ and $w$. The top left plot has $b=1/3,2/3,1,2,3$ in order of increasing dash width ($w=-1$), the bottom plot has $w=-1,-0.75,-0.5,-0.4,-0.35$ ($b=1$) in order of increasing dash width. In the middle column, the top plot has $q=0,0.25,0.5,0.75,1$ as the dash width increases ($w=-1$), and $w$ varies as its lefthand neighbor in the bottom plot ($q=0.5$). Finally, in order of increasing dash width, the top right plot has $b=1/3,2/3,1,2,3$ ($q=0.5$) and the bottom right plot has  $q=0,0.25,0.5,0.75,1$ ($b=1$).}
\label{plot04}
\end{figure*}

The probability that observers measure $r$ at its present value or larger is then given by:
\begin{align}
P(r\geq0.4)&=\int_{0.4}^1 p(r)\textrm{d}r\, .
\end{align}
In the case that $q=1/2$, $b=1$ and $w=-1$, this evaluates to the far-from-fine-tuned $P(r\geq0.4)=0.71$, and this result is relatively insensitive to the particular values of $b$,  $q$ and $w$. Plots of $P(r\geq0.4)$ as a function of $b$, $q$ and $w$ are given in figure \ref{plot04}, where it can clearly been seen that for all plotted values the probability remains between 0.4 and 1, and so the absence of fine-tuning is robust to varying these parameters .

Before moving on to a discussion of the above results, I'd like to briefly return to an observation from the beginning of this section. So far only the number modes has been considered as constraining the ability of an observer to do cosmology. However, as well as the number of measurements, one might also consider the ease of making those measurements. The latter quantity is not as easy to find a suitable proxy for, but one possibility could be the ability of cosmologists to distinguish between accelerating and non-accelerating cosmologies. This could entail, for example, finding an expression for how the difference in the luminosity-redshift relation between an accelerating and decelerating cosmology varies as a function of $r$. If one did this, one would find a slightly greater pressure towards large $r$ when $\Omega_m>\Omega_{\Lambda}$ and towards small $r$ when $\Omega_m<\Omega_{\Lambda}$. However, numerical investigations suggest that the probabilities are altered by around $10\%$, which would have no effect on the above conclusions. Moreover, in the absence of a compelling reason to do so, it is better to model selection effects with as little sensitivity to detailed physics as possible.

\section{Caveats and Conclusions}\label{conclusions}
The strengths and weaknesses of the above argument are reviewed below, but before we get to those there are a couple of as yet undiscussed caveats that should be mentioned. The first is that the reasoning used herein does not apply to dynamical dark energy models, where the period of acceleration is temporary. Of course, this simply means that the ``why now?'' problem must be added to the list of challenges such models face. In addition, by not including $\Omega_r$ in my model, I have implicitly ignored the (possible) coincidence of our epoch of existence and that of matter-radiation equality. Such a coincidence is considerably milder in magnitude than that of the dark energy and matter coincidence, with $\Omega_r/\Omega_m\sim10^{-4}$. However, the naive expectation should be that this ratio is close to its lowest possible value. There may well be an anthropic explanation for this fine(ish)-tuning, but in this paper it remains a mystery.

The calculations in the previous section demonstrate that the coincidence problem is an artifact of selection bias. This demonstration required the following assumptions:
\begin{itemize}
\item Selection effects are a sufficient explanation of fine-tuning.
\item An expectation that we are more likely to find ourselves measuring $r$ where most of the measurements of $r$ are possible is a sensible selection effect.
\item The frequency of measurements of $r$ is correlated with the number of cosmological modes within a single Hubble radius, $N(r)$.
\item There are no other factors that have a significant effect on the frequency of measurements of $r$.
\end{itemize}
With these assumptions the argument follows straightforwardly: observers are more likely to measure $r$ when it's easy to measure, $r$ is easy to measure when there are lots of modes available to the observer, there are lots of modes available to the observer when $r$ is close to 1, q.e.d. Of course, there still remains the justification of the above assumptions. While the above points have been defended at the relevant points in the body of the text, it is useful to review the arguments before we finish.

Defenses of the anthropic principle are manyfold, and there is little I can offer that will persuade the unpersuaded reader. That said, I suppose it behoves one to try. Fine tuning problems can, in the most part, be viewed as statements about selection effects, if not in real space at least in the space of possible worlds. This is especially true with regards to the cosmological coincidence, where the problem can be rephrased as: ``If our epoch of existence is selected (log) uniformly in time, why are we so fortunate as to live in the epoch of matter and dark energy equality?'' All the principle of mediocrity states is that existence is not selected from a uniform distribution and that we can make reasonable deductions about what that distribution should be.

Of course, making ``reasonable'' deductions, is far from straightforward. In the case of this paper I have argued that cosmologists should expect to find themselves living in the epoch when most cosmology can be done, and that furthermore, this epoch is the one in which the greatest number of visible modes. The correlation between number of modes and number of measurements contains an implicit assumption that there is bound structure. Moreover, there are methods (though somewhat constrained ones) to continue cosmology after all else but the local structure has exited the horizon; one such method is the measurement of cosmological parameters using hyper-velocity stars, discussed by Loeb in  \cite{Loeb:2011cw}. With that said, precision cosmology will be certainly be harder in the future, even if it is not impossible.

The use of only $N(r)$ to construct $p(r)$ can be defended on several grounds. Firstly, incorporating additional dependence on the ease of discriminating between cosmologies did not substantially change my conclusions. Secondly, the peakedness of $N(r)$ suggests that it would take a substantial anthropic counterweight to restore a pressure to small values of $r$, and one has trouble conceiving of what such an effect could be. Finally, the calculation of $N(r)$ requires little additional physics and is mostly insensitive to additional cosmological parameters (in particular, details of the spectrum, of structure formation, of the physics of radiation and so on); this suggests that marginalizing over additional parameters would not effect the form of $N(r)$.

The alert reader will have realized that I glossed over a key point in my defense of the assumptions of this paper: Why should cosmologists expect to live when cosmology is easiest? Isn't my whole argument trivial? Really all that's been done is to show that we live in the era of engaging cosmology, without answering the question of \emph{why} we live in interesting times.

Well, in much the same way as students of climate change were unlikely to be found before the climate started changing, so it is with cosmologists. In epochs where cosmology is verging on the impossible, the questions about the apparent interestingness (or otherwise) of cosmology are unlikely to be asked.

We are fortunate enough to live in interesting times, but if we did not, we would be blissfully unaware of that fact.

\begin{acknowledgments}
I would like to thank Bruce Bassett, Rhiannon Gwyn and Michael Salem for helpful discussions.
\end{acknowledgments}

\begin{widetext}
\end{widetext}

\appendix
\section{$p(r)$ Without Selection Effects}\label{withoutselection}
Let's begin by considering $p(r)$ when $p(a)\sim1$
\begin{align}
p(r)&=\frac{\textrm{d}a}{\textrm{d}r}p(a)\sim\frac{\textrm{d}a}{\textrm{d}r}\, .
\end{align}
Noting $r\sim a^3$ when $\Omega_m>\Omega_{\Lambda}$ and $r\sim a^{-3}$ when $\Omega_{\Lambda}>\Omega_m$, this gives:
\begin{align}
p\left(r|\Omega_m>\Omega_{\Lambda}\right)&\sim r^{-2/3}\nonumber\\
p\left(r|\Omega_{\Lambda}>\Omega_m\right)&\sim r^{-4/3}\, .
\end{align}
To calculate $p(r)$ we also need $P(\Omega_m>\Omega_{\Lambda})$ and $P(\Omega_{\Lambda}>\Omega_m)$:
\begin{align}
P(\Omega_m>\Omega_{\Lambda})&=\frac{\int_{a_i}^{a_1}\textrm{d} a}{\int_{a_i}^{a_f}\textrm{d} a}=\frac{a_1-a_i}{a_f-a_i}\sim0\\
P(\Omega_{\Lambda}>\Omega_m)&=\frac{\int_{a_1}^{a_f}\textrm{d} a}{\int_{a_i}^{a_f}\textrm{d} a}=\frac{a_f-a_1}{a_f-a_i}\sim1\, .
\end{align}
$a_1$ is the value of $a$ when $r=1$. The last approximate equality in each line corresponds to taking the limit $a_i\ll a_1\ll a_f$. Thus:
\begin{align}
P(r\geq0.4)&=\frac{\int_{0.4}^1r^{-4/3}\textrm{d}r}{\int_{r_f}^1r^{-4/3}\textrm{d}r}\nonumber\\
&=\frac{1.07}{r_f^{-1/3}-1}\sim\frac{a_1}{a_f}\sim e^{-Ht_f}\, .
\end{align}
$H$ is the asymptotic value of the Hubble constant for a cosmological constant dominated universe, $H^2=\Lambda/3$. A cutoff ($r_f$, $t_f$, $a_f$) is introduced in the normalization of the probability to keep everything finite. As a result, one finds that, because of the large future volume, the probability of measuring $r\geq0.4$ is exponentially small as a function of the cutoff time.

Now let us consider $p(r)$ when $p(\textrm{ln\,} a)\sim1$. With the dependence of $r$ on $a$ as before, this gives:
\begin{align}
p\left(r|\Omega_m>\Omega_{\Lambda}\right)&\sim r^{-1}\, ,&p\left(r|\Omega_{\Lambda}>\Omega_m\right)&\sim r^{-1}\, .
\end{align}
Since $p(a)\sim a^{-1}$,  $P(\Omega_m>\Omega_{\Lambda})$ and $P(\Omega_{\Lambda}>\Omega_m)$ are given by:
\begin{align*}
P(\Omega_m>\Omega_{\Lambda})&\sim\frac{\ln\frac{a_1}{a_i}}{\ln\frac{a_f}{a_i}}\, ,&P(\Omega_{\Lambda}>\Omega_m)&\sim\frac{\ln\frac{a_f}{a_1}}{\ln\frac{a_f}{a_i}}\, .
\end{align*}
Then for the probability that $r\geq0.4$, one has:
\begin{align}
P(r\geq0.4)&=\frac{1}{\ln\frac{a_f}{a_i}}\left(\int_{0.4}^1r^{-1}\textrm{d}r\right)\left(\frac{\ln\frac{a_1}{a_i}}{-\ln r_i}-\frac{\ln\frac{a_f}{a_1}}{\ln r_f}\right)\nonumber\\
&\sim\frac{1}{\ln\frac{a_f}{a_i}}\left(\frac{\ln\frac{a_1}{a_i}}{-3\ln\frac{a_i}{a_1}}-\frac{\ln\frac{a_f}{a_1}}{-3\ln\frac{a_f}{a_1}}\right)\nonumber\\
&\sim\frac{1}{\ln\frac{a_f}{a_i}}\sim\frac{1}{Ht_f-\frac{2}{3}\ln\frac{t_i}{t_1}}\, .
\end{align}
Once again the probability of measuring $r\sim\mathcal{O}(1)$ is determined by the cutoffs. In this case both early and late time cutoffs are important, though the sensitivity is somewhat less.

\section{A Different Choice for $f_i$}\label{threshold}
Instead of taking a power law relationship between $p(r)$ and $N(r)$, one can take a threshold approach, where all measurements are considered equal when the number of modes available to the observer is greater than some minimum value.

To calculate the value of $P(r\geq0.4)$ in such a model of measurement requires more than just specifying a minimum number of modes; one also needs to define ``equally likely''. There are a number of different interpretations of the assumption that measurements are equally probable after the threshold number of modes has been exceeded. Three reasonable possibilities are $p(r)=\textrm{const}$, $p(a)=\textrm{const}$ and $p(\ln a)=\textrm{const}$. These will give different values of the amount (or otherwise) of fine-tuning inherent in our measurements of $r$.

Considering the $w=-1$ case and setting the minimum number of modes needed to observe $r$ as $n$, one finds:
\begin{align}
p(r)&=\textrm{const}\, , & P(r\geq0.4)&\sim\frac{0.6}{1-r_0n^2}\, ,\nonumber\\
p(a)&=\textrm{const}\, , & P(r\geq0.4)&\sim(r_0^{1/2}n)^{1/3}\, ,\nonumber\\
p(\ln a)&=\textrm{const}\, , & P(r\geq0.4)&\sim\frac{1}{-\ln\left[r_0^{1/2}n\right]}\, .
\end{align}
The quantity $r_0^{-1/2}$ is approximately equal to the maximum number of modes that will ever available to an observer, so $r_0^{1/2}n$ is the fraction of the maximal data available that is needed to make an observation. In the first case above $P(r\geq0.4)$ is not finely tuned at all. In the second and third cases the tuning is worst if we assume $n=1$. For our universe, where $r_0\sim10^{-12}$, this is a fine-tuning of $\sim1/100$ for $p(a)=\textrm{const}$ and $\sim1/10$ for $p(\ln a)=\textrm{const}$.

\bibliography{coincidence}

\end{document}